\documentclass[runningheads]{llncs}
\usepackage{graphicx}
\usepackage{amsmath}
\usepackage{amssymb}
\usepackage{booktabs}
\usepackage{color}
\usepackage[misc]{ifsym}
\usepackage[symbol]{footmisc}
\begin{document}
\title{Learning Hybrid Representations for Automatic 3D Vessel Centerline Extraction}
\titlerunning{Hybrid Representations for Vessel Centerline Extraction}
%

\author{Jiafa He\inst{1\ast} \and 
Chengwei Pan\inst{2}\thanks{These authors contributed equally to this work.} \and 
Can Yang\inst{1} \and Ming Zhang\inst{2} \and Yang Wang\inst{1} \and \\ 
Xiaowei Zhou\inst{3}\textsuperscript{$\dagger$}$^{\left(\textrm{\Letter}\right)}$ \and
Yizhou Yu\inst{4}\thanks{Corresponding authors. Email: xzhou@cad.zju.edu.cn, yizhouy@acm.org}$^{\left(\textrm{\Letter}\right)}$
}
%
\authorrunning{J. He \it{et al.}}
%
\institute{\textsuperscript{1}Department of Mathematics, The Hong Kong University of Science and Technology, Hong Kong, China\\
\textsuperscript{2}Department of Computer Science, School of EECS, Peking University, Beijing, China \\
\textsuperscript{3}The State Key Lab of CAD\&CG, Zhejiang Unversity, Hangzhou, China \\
\textsuperscript{4}Deepwise AI Lab, Beijing, China\\}
\maketitle                
\begin{abstract}
Automatic blood vessel extraction from 3D medical images is crucial for vascular disease diagnoses. Existing methods based on convolutional neural networks (CNNs) may suffer from discontinuities of extracted vessels when segmenting such thin tubular structures from 3D images. We argue that preserving the continuity of extracted vessels requires to take into account the global geometry. However, 3D convolutions are computationally inefficient, which prohibits the 3D CNNs from sufficiently large receptive fields to capture the global cues in the entire image.
In this work, we propose a hybrid representation learning approach to address this challenge. The main idea is to use CNNs to learn local appearances of vessels in image crops while using another point-cloud network to learn the global geometry of vessels in the entire image. In inference, the proposed approach extracts local segments of vessels using CNNs, classifies each segment based on global geometry using the point-cloud network, and finally connects all the segments that belong to the same vessel using the shortest-path algorithm. This combination results in an efficient, fully-automatic and template-free approach to centerline extraction from 3D images. We validate the proposed approach on CTA datasets and demonstrate its superior performance compared to both traditional and CNN-based baselines.

\end{abstract}
\section{Introduction}
Extracting tubular objects, e.g., blood vessels, has become a crucial task in computer-assisted diagnosis (CAD) of many diseases. For example, vessel lumen segmentation and centerline extraction are prerequisites for vessel curved-planar reconstruction (CPR)~\cite{kanitsar2002cpr} from computed tomography angiography (CTA) images, which further facilitates stenosis detection and plaque identification in clinical diagnosis. However, it is usually time-consuming to segment vessel and extract centerline from various medical images. Instead, automatic vessel segmentation and centerline detection play a more and more important role for quantitative analysis of vascular diseases.~\cite{cetin2015higher}~\cite{kitamura2012automatic}~\cite{zhai2019automatic}.

Recently, convolutional neural networks (CNNs) have been widely applied in 3D medical image segmentation. However, segmenting vessels in 3D medical images is still very challenging. The blood vessels have delicate tubular structures with a large variety in long-range topology, which cannot be captured by slice-wise or patch-wise convolutional operations in most deep learning based segmentation methods~\cite{kamnitsas2017efficient}~\cite{litjens2017survey}~\cite{yu2018recurrent}. Moreover, at the presence of imaging artifacts which often exist in medical images, CNN-based segmentation algorithms are prone to missing some segments of vessels, resulting discontinuities in the extracted vessel centerline~\cite{stefancik2001highly}~\cite{yang2012automatic}~\cite{zheng2013robust}. To preserve the correct topology of the extracted vessels, previous methods may rely on user input that annotates start and end points of each vessel~\cite{selvan2018extracting}~\cite{wolterink2019coronary}. Then, a more complete vessel centerline can be found by a minimal-cost path-based algorithm~\cite{deschamps2001fast}~\cite{gulsun2016coronary}~\cite{li2007vessels}~\cite{moriconi2017vtrails}. To avoid manual input, another approach to ensure correct topology is to build an atlas or template of the target vessels from training samples and register the template to the test image~\cite{zhang2010coronary}~\cite{zheng2013robust}. However, this approach is not very generalizable as the vascular structure of the test sample might be very different from the template.

In this paper, we present a novel approach for vessel centerline extraction, which is able to ensure the connectivity of extracted vessels without the need of any manual input or vessel template. The key idea is to use a patch-wise 3D CNNs to segment vessel mask and regress vessel centerline heatmap from the input image, and meanwhile use another point-cloud network to label the extracted vessel segments, such that segments belonging to the same vessel can be connected in a post-processing step. This hybrid approach makes the best use of both worlds: patch-wise CNNs for local appearance learning and point-cloud networks for global geometry learning, resulting in a robust and efficient algorithm for automatic centerline extraction. We also propose a geometry-aware grouping strategy to improve the performance of point-cloud network for vessel labeling. The effectiveness of the proposed framework is validated on two datasets: a public dataset of coronary artery CTA scans and an in-house dataset of head and neck artery CTA scans. Experimental results show that our approach outperforms existing baseline methods in terms of both accuracy and completeness of extracted centerlines.

In summary, we make the following contributions: (1) A novel hybrid representation learning approach for fully-automatic and template-free vessel centerline extraction;
(2) A geometry-aware grouping method that utilizes the skeleton's connection property to improve the performance of vessel labeling;
(3) The state-of-the-art performance on the public benchmark.
\section{Methods}

Given a 3D CTA image consisted of a sequence of 2D slices, the objective is to segment the arteries and delineate their centerlines. The state-of-the-art segmentation methods are mostly based on CNNs. Due to the heavy computation of 3D convolutions, an input 3D image needs to be divided into overlapped patches and fed into the segmentation network separately. This leads to restricted local receptive fields, which may not provide sufficient information to distinguish between arteries and veins, resulting false detection. Moreover, there is no guarantee on the connectivity of extracted vessel segments from patch-based CNNs. A solution is to connect the segments that belong to the same vascular branch in post-processing. But to achieve this we need to label all the segments, which is also difficult for a patch-based CNN as vessel labeling requires considering the global geometry of the vessels.

\begin{figure*}[t]
\begin{center}
\includegraphics[width=\linewidth]{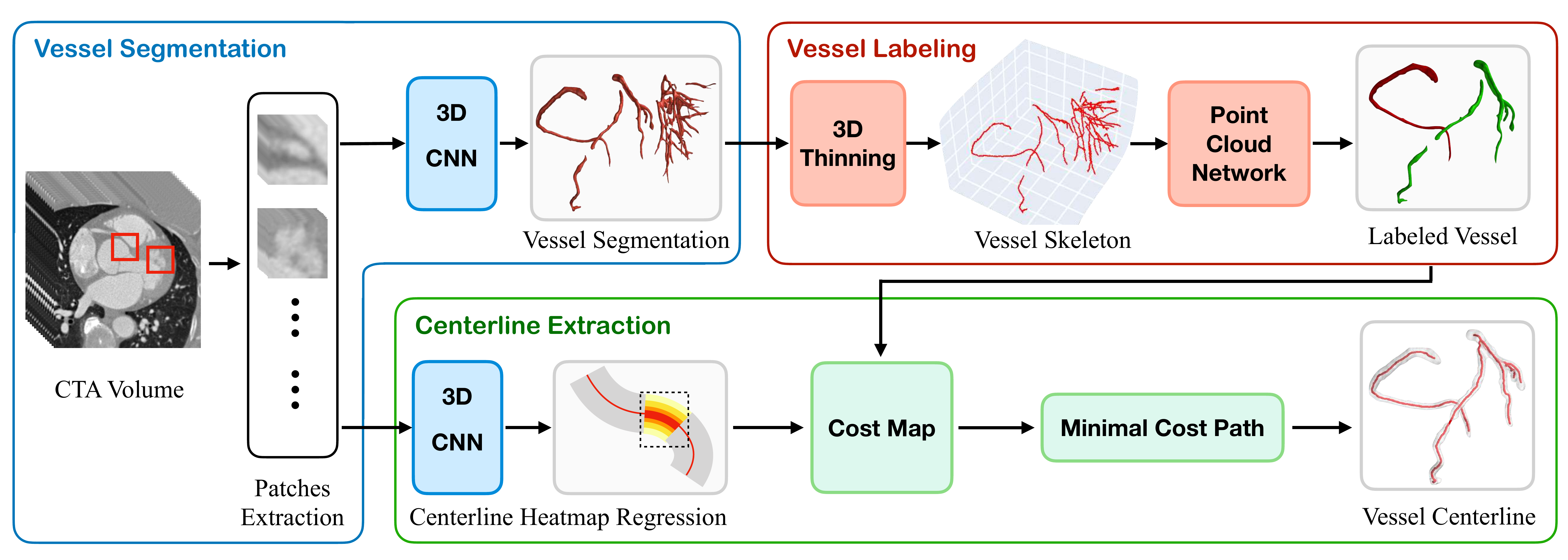}
\end{center}
  \caption{\textbf{Overview of the proposed approach}. The framework consists of three components: \textbf{Vessel segmentation ({blue block})}: the input CTA images are divided into 3D patches with overlap and then a 3D CNNs are utilized to efficiently learn local features of vascular objects. \textbf{Vessel labeling {(red block)}}: vascular skeletons produced by thinning the segmentation results are fed into a point-cloud network to learn the global geometry of vessels and realize vascular branch labeling. \textbf{Centerline extraction {(green block)}}: based on a cost map, which is constructed from the centerline heatmap and the labeled skeleton, a minimal cost path algorithm is finally utilized to extract the complete vessel centerline.}
\label{fig2}
\end{figure*}

To address these issues, we propose a hybrid approach that consists of a patch-based CNNs for local \textbf{vessel segmentation}, a point-cloud network for global \textbf{vessel labeling} and a path-finding algorithm for final \textbf{centerline extraction}. Figure \ref{fig2} provides an overview to our approach.

\subsection{Vessel Segmentation}\label{vessel segmentation}
At first, we use a 3D CNNs to learn vascular local appearance features from 3D patches of the original CTA images and produce a coarse vessel segmentation. An UNet~\cite{ronneberger2015u} backbone architecture with an encoding-decoding module is selected to transform the input image to the segmentation mask. Moreover, to explore long-range contextual information inside 3D patches, we embed a dual attention module~\cite{fu2019dual} on top of the UNet backbone. Finally, a combination of a binary cross-entropy loss and a Dice loss is used as the total segmentation loss. Please refer to the supplementary for more details.

\subsection{Vessel Labeling}\label{Vessel Labeling}
\begin{figure}[t]
\begin{minipage}[t]{0.48\linewidth}
\centering
\includegraphics[width=1.0\textwidth]{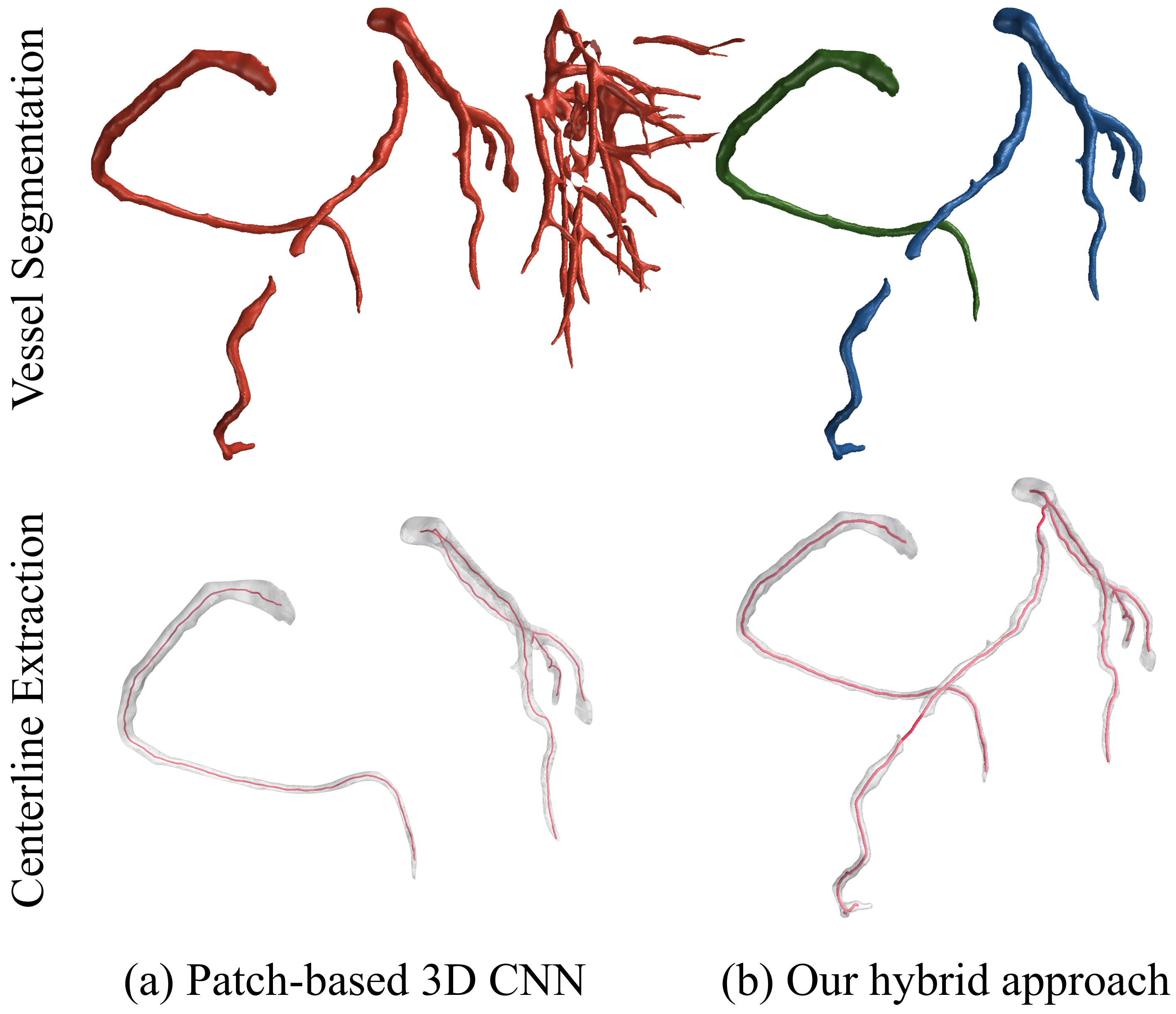}
\caption{\textbf{Comparison}. Patch-based 3D CNNs extract extra non-vascular tissues and miss some true segments (left). Our hybrid approach is able to remove non-vascular tissues and connect the disjointed segments (right). \label{fig1}}
\end{minipage}
\hfill
\begin{minipage}[t]{0.50\linewidth}
\centering
\includegraphics[width=1.0\textwidth]{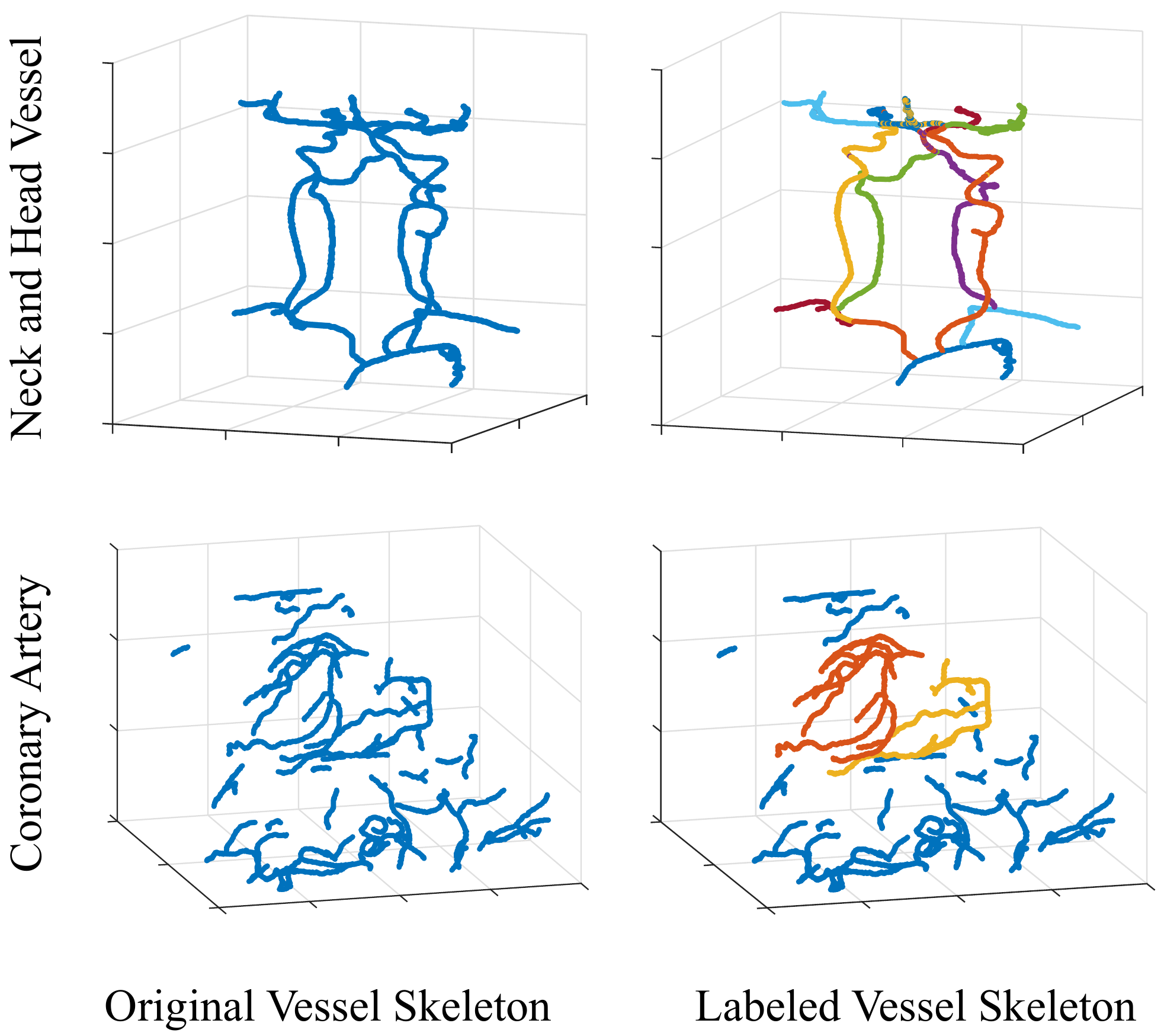}
\caption{\textbf{Generated vessel skeletons}. Head and neck arteries (the top row) are divided into 17 categories distinguished by color. Coronary artery (the bottom row) are divided into 3 categories including a category for false positives.\label{fig3}}
\end{minipage}
\end{figure}

Due to the lack of the global information in the patch-wise 3D CNNs, the vessel segments obtained from the vessel segmentation procedure tend to contain some false-positive results like veins and also miss some parts of tiny or tortuous vessels.
We propose to perform a vessel labeling procedure that classifies the segmented vessels into different branches. Then such semantic labels can be used to remove non-vascular segments and group discontinuous vessel segments, as shown in Figure \ref{fig1}. The vessel labeling procedure is implemented by first generating a set of points which represent vascular skeleton from the vessel segmentation results and then using a point-cloud network to predict labels of these skeleton points.

\noindent{\bf Point-cloud Generation.}
As it is inefficient to directly label vessel segments in 3D volumes using CNNs, we propose to perform the labeling on vascular skeletons represented by a set of points, which can reduce complexity and preserve original geometric information of vessels.
To generate vascular skeletons from the vessel segments, a 3D thinning algorithm~\cite{palagyi2001sequential} is applied to erode the vessel segments and finally obtain single-voxel-width skeleton points, as shown in Figure \ref{fig3}. The generated skeletons are composed of discrete and unordered points lying on the center of the vascular lumen, which represent vascular geometry.

\noindent{\bf Vessel Labeling Network.}
Given a set of vascular skeleton points generated from vessel segments ${\{Q_{i}\in\mathbb{R}^{3}, i\in 1,2,...,M\}}$, the target of the vessel labeling network is to predict the label of each point, ${f(Q_{i})\rightarrow Y_{j}\,(Y_{j}\in\mathbb{Z}, j\in 1,2,...,K)}$, as shown in Figure \ref{fig3}. It is similar to point-cloud semantic segmentation tasks in 3D computer vision.
Any point-cloud network can be adopted, such as the state-of-the-art PointNet++~\cite{qi2017pointnet++} and dynamic graph CNNs (DGCNN)~\cite{wang2018dynamic}. We will provide a comparison between them in experiments.

\begin{figure}[t]
\begin{minipage}[t]{0.46\linewidth}
\centering
\includegraphics[width=0.8\textwidth]{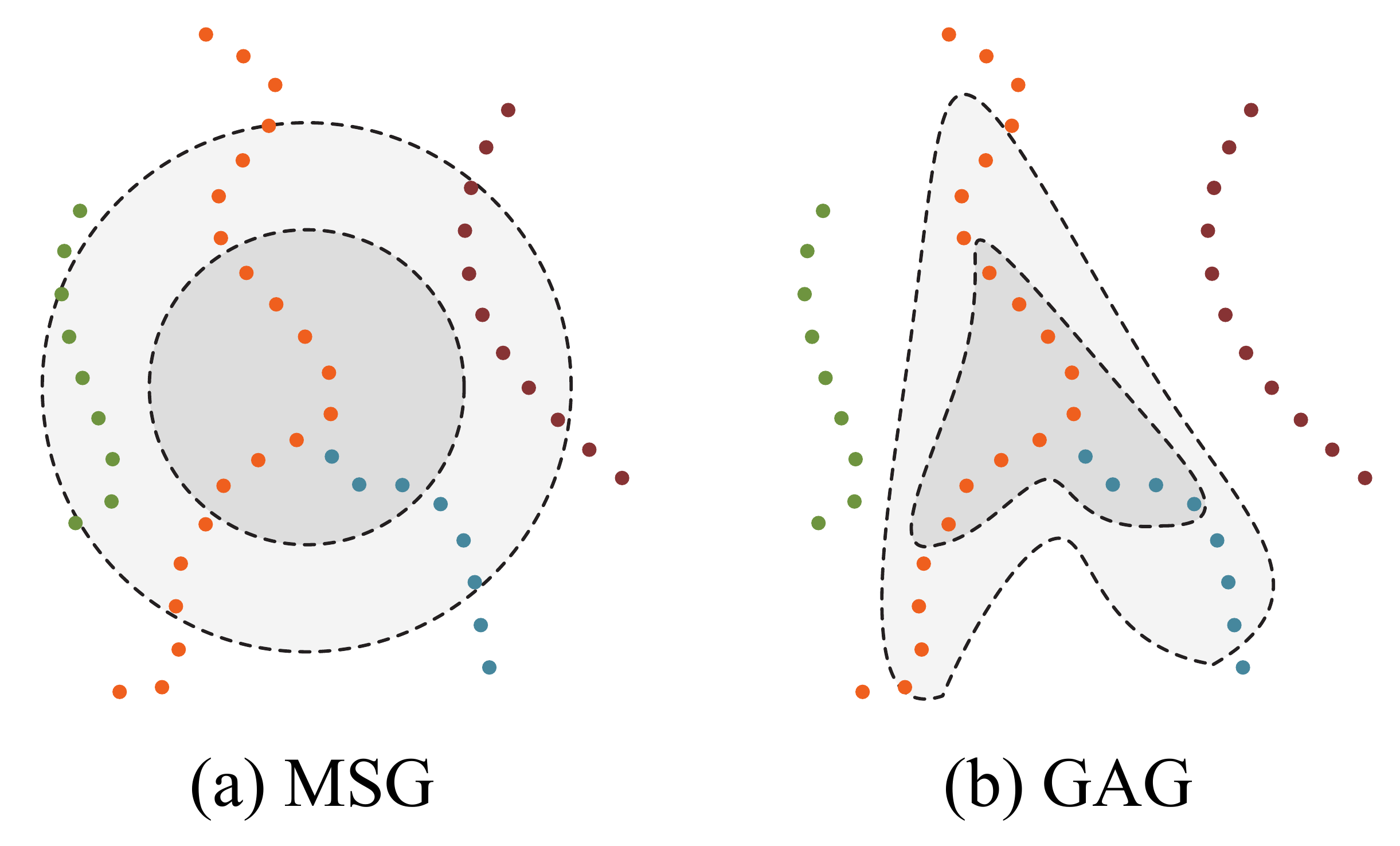}
\caption{\textbf{Comparison of points grouping methods.}  (a) \textbf{Multi-scale grouping (MSG) in the PointNet++}: the grouping areas are multi-scale $\epsilon$-sphere and grouped points may come from different skeleton components. (b) \textbf{Geometry-aware grouping (GAG)}: the grouping areas trend to stretch along the lines and group points from the same islands.
\label{fig5}}
\end{minipage}
\hfill
\begin{minipage}[t]{0.52\linewidth}
\centering
\includegraphics[width=1.0\textwidth]{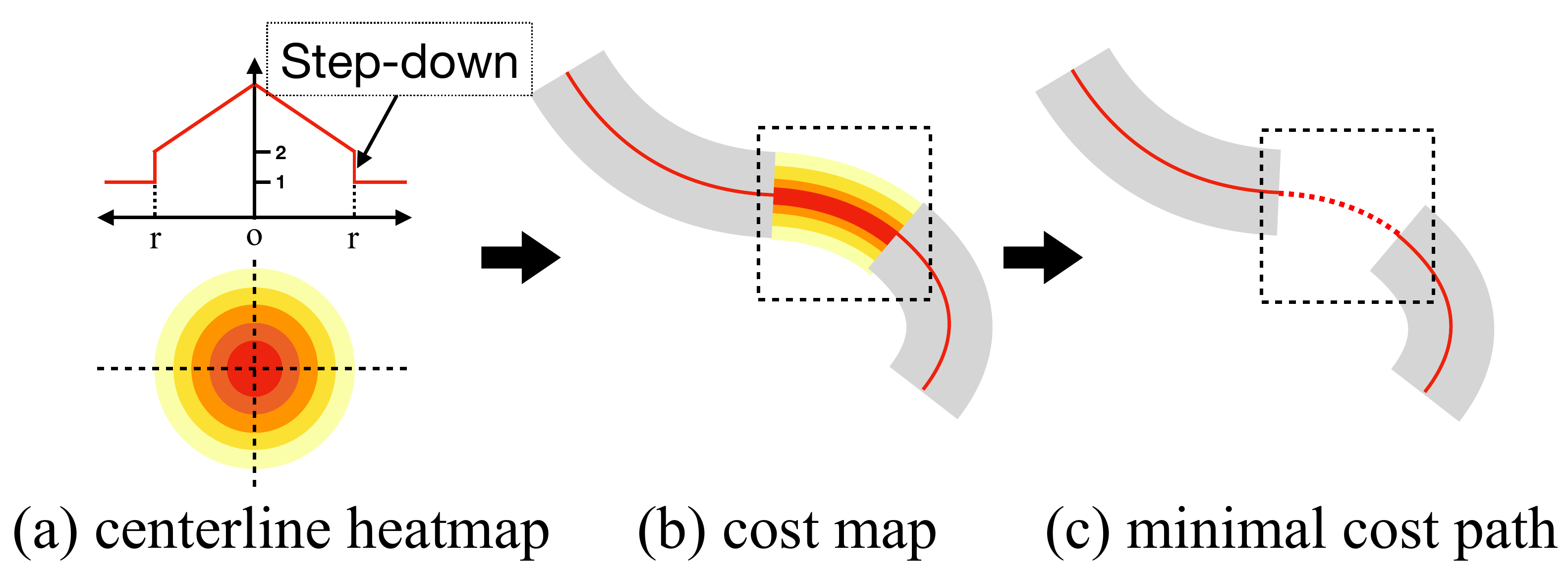}
\caption{\textbf{Centerline Extraction.} (a) \textbf{Centerline heatmap}: the heatmap has larger value in point closer to the centerline and the points laying outside the vascular radius are set to a step-down low value. (b) \textbf{Cost map}: the cost map is built by combining the heatmap and the skeleton lines. (c) \textbf{Minimal cost path}: based on the cost map, a minimal cost path can be extracted to connect the broken lines as the complete vessel centerline. \label{fig6}}
\end{minipage}
\end{figure}

\noindent{\bf Geometry-aware Grouping.}
A particular property of the skeleton points compared to a general point cloud is the given connectivity among adjacent skeleton points.
Specifically, skeleton points can be divided into separated components using a connected-component labeling (CCL) algorithm~\cite{park2000fast}.
However, both PointNet++ and DGCNN are realized to group points by $k$-nearest neighbor ($k$-NN) algorithm based on L2 distance, which ignores the given geometry of the vessel skeletons. As demonstrated in Figure \ref{fig5}(a), the L2-based methods are more likely to group points belonging to different components. To address this issue and leverage the skeletons' connection property, we propose a geometry-aware grouping method (GAG) to modify the distance based on the connection relationship. The modified distance between two points is computed by
\begin{equation}
\begin{aligned}
\mathbf{D}_{i,j} = \begin{cases}
 \lambda_{n} \left \| \mathbf{x}_{i}-\mathbf{x}_{j} \right \|_{2} &\text{ if } \mathbf{x}_{i}, \mathbf{x}_{j} \in C_{n}. \\
 (1-\lambda_{n}) \left \| \mathbf{x}_{i}-\mathbf{x}_{j} \right \|_{2}& \text{otherwise.}
\end{cases}
\end{aligned}
\end{equation}
\noindent where $\lambda_{n}$  $(\lambda_{n}<0.5)$ is a weight, $C_{n}$ is the n-th component. As shown in Figure \ref{fig5}(b), the grouping area in GAG is prone to stretching along the skeleton lines and the points belonging to the same connected component are more likely to be grouped. This design facilitates local feature consistency in the same component and consequently improves the accuracy of vessel labeling as evaluated in the experiments.

\subsection{Centerline Extraction}\label{Centerline Extraction}
After vessel labeling, each vessel segment is assigned a semantic label. The semantic label can be used to remove non-vascular tissues and provide the guidance to connect disjointed vessel segments. Specifically, we first determine which segments should be connected and then use a minimal cost path method to connect them and output final centerlines based on a cost map. The cost map is constructed by a centerline heatmap regressed from another 3D CNNs and the labeled vessel skeletons.

\noindent{\bf Centerline Heatmap Regression.}
The centerline heatmap is defined as the opposite of a distance map, where points closer to the vessel centerline have larger values and points outside the vascular radius have a step-down low value, as shown in Figure \ref{fig6}(a). Considering that the probability map obtained from the segmentation network has only learned the difference between background and the vessel (e.g. edge features) rather than the centerline feature within the vessel lumen, we adopt another 3D CNN similar to the network used in vessel segmentation to regress the centerline heatmap. The mean square error (MSE) loss is used to train the network. Based on the centerline heatmap, we construct a cost map to guide the minimal cost path search algorithm as shown in Figure \ref{fig6}(b).
In areas where vessel segments exist, the cost map directly assigns a large value to the skeleton points (red lines in Figure \ref{fig6}(b)) and a small value to the points elsewhere (gray areas in Figure \ref{fig6}(b)).

\noindent{\bf Minimal Cost Path.}
Given vessel skeletons and their labels, we successively merge the disjointed segments with the same label. Paired boundary points with the same label are put in a priority queue according to the distance  between  the  two  points  in  the  pair. We take a pair successively from the queue and then the minimal cost path is found by Dijkstra algorithm~\cite{verscheure2010dijkstra} to connect the two boundary points, as shown in Figure \ref{fig6}(c). If the two segments represented by the two points are connected in the previous step, we skip the pair and go on until the queue is empty.

\begin{table}[t]
\renewcommand\arraystretch{1.1}
\caption{Performance comparison of automatic coronary artery centerline extraction methods. OV represents the completeness of extracted vessel centerline and is similar to Dice coefficient. OF determines the ratio of a coronary artery that has been extracted before making an error. OT gives an indication of how much the extracted centerline overlaps with the clinically relevant centerline reference (radius $\geq$ 0.75 mm).}\smallskip
\centering
\resizebox{.6\columnwidth}{!}{
\smallskip\begin{tabular}{cccc}
\toprule
Methods & OV(\%) & OF(\%) & OT(\%) \\
\hline
ModelDrivenCenterline~\cite{zheng2013robust} & 92.4 & 80.6 & 93.4\\
SupervisedExtraction~\cite{kitamura2012automatic} & 90.6 & 70.9 & 92.5\\
GFVCoronaryExtractor~\cite{yang2012automatic} & 93.7 & 74.2 & 95.9\\
DepthFirstModelFit~\cite{zambal2008shape} & 84.7 & 65.3 & 87.0\\
AutoCoronaryTree~\cite{tek2008automatic} & 84.7 & 59.5 & 86.2\\
\hline
Our approach & \textbf{95.8} & 72.3 & \textbf{96.3}\\
\bottomrule
\end{tabular}
}
\label{table2}
\end{table}

\begin{figure}[htb]
\begin{minipage}[t]{0.50\linewidth}
\centering
\includegraphics[width=1.0\textwidth]{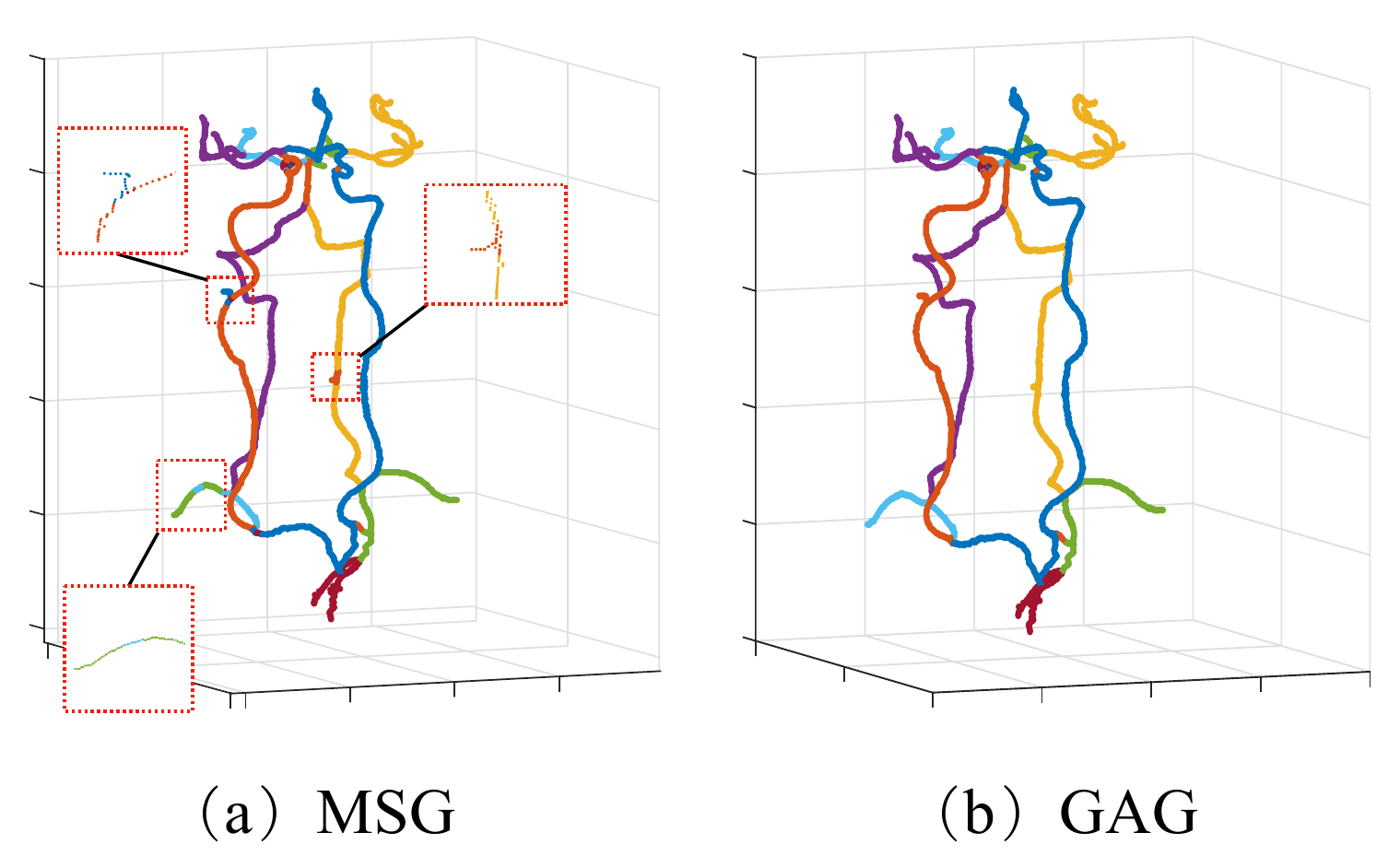}
\caption{Comparison of different grouping methods used in vessel labeling networks. It can be seen that the proposed GAG method achieves better local consistency of the predicted vessel labels.
\label{fig7}}
\end{minipage}
\hfill
\begin{minipage}[t]{0.48\linewidth}
\centering
\includegraphics[width=1.0\textwidth]{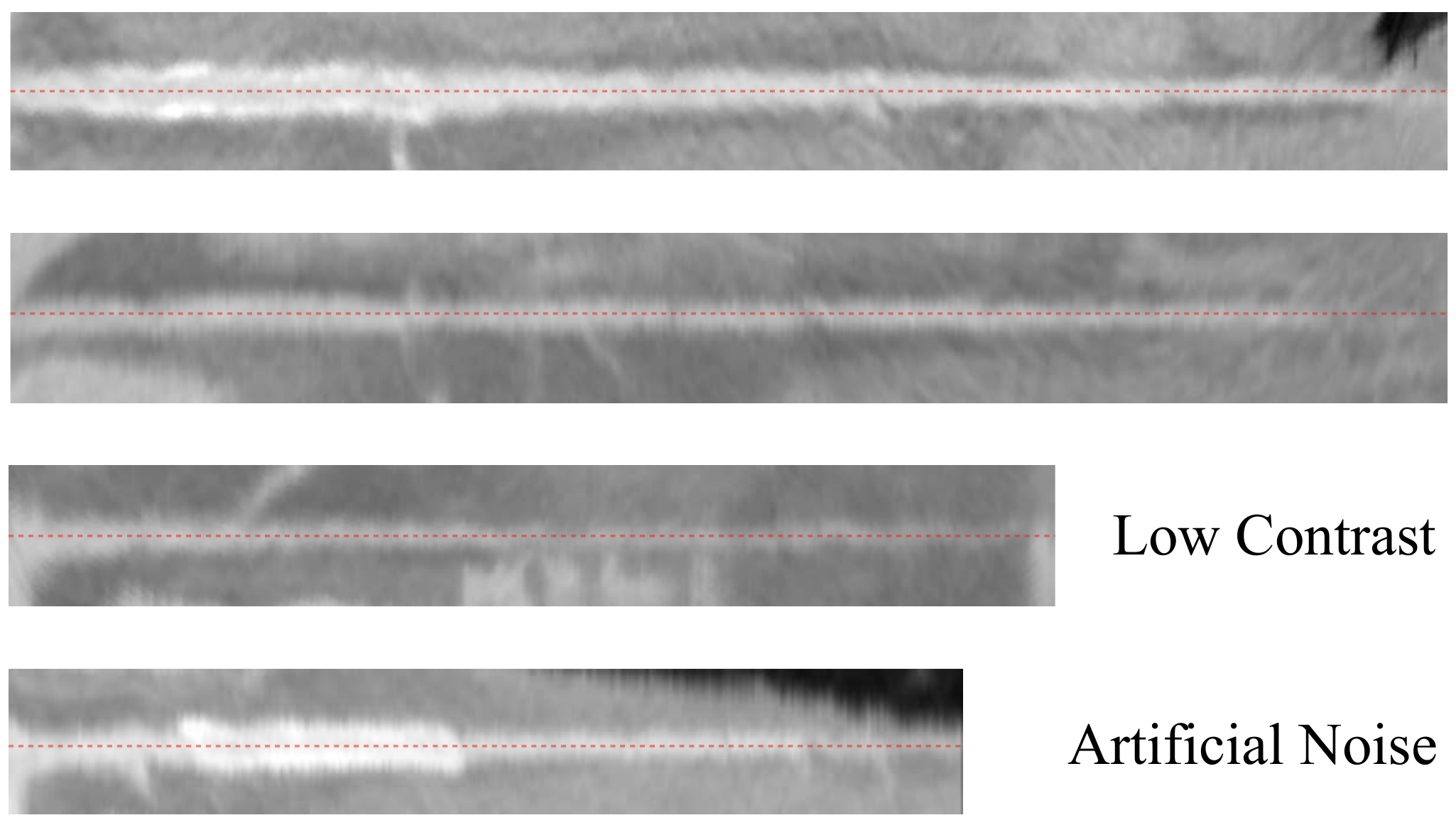}
\caption{Vessel curved-planar reconstruction (CPR) of the coronary artery based on the extracted centerline. Our approach can extract complete centerlines under serious imaging artifacts.
\label{fig8}}
\end{minipage}
\end{figure}

\begin{table}[t]
\renewcommand\arraystretch{1.1}
\caption{Performance comparison of different variants of vessel labeling networks on the private head and neck artery dataset. Four metrics are reported: the accuracy of vessel labeling and OV, OF and OT of four main vessel centerlines including left  and  right  common  carotid  artery  (L/RCCA),  left and right vertebral artery (L/RVA).
}\smallskip
\centering
\resizebox{.6\columnwidth}{!}{
\smallskip\begin{tabular}{ccccc}
\toprule
Vessel Labeling & Accuracy(\%) & OV(\%) & OF(\%) & OT(\%) \\
\hline
PointNet~\cite{qi2017pointnet} & 92.2 & 96.1 & 84.5 & 96.6\\
PointNet++~\cite{qi2017pointnet++}  & 95.1 & 96.5 & \textbf{85.8} & 97.0\\
PointNet++(GAG) & 97.0 & \textbf{97.2} & 85.7 & \textbf{97.6}\\
DGCNN~\cite{wang2018dynamic} & 96.5 & 96.6 & 84.4 & 97.1 \\
DGCNN(GAG) & \textbf{97.2} & 97.0 & 84.9 & 97.4\\
\bottomrule
\end{tabular}
}
\label{table4}
\end{table}

\begin{figure*}[htb]
\setlength{\abovecaptionskip}{0.cm}
\setlength{\belowcaptionskip}{-0.cm}
\centering
\includegraphics[width=0.9\linewidth]{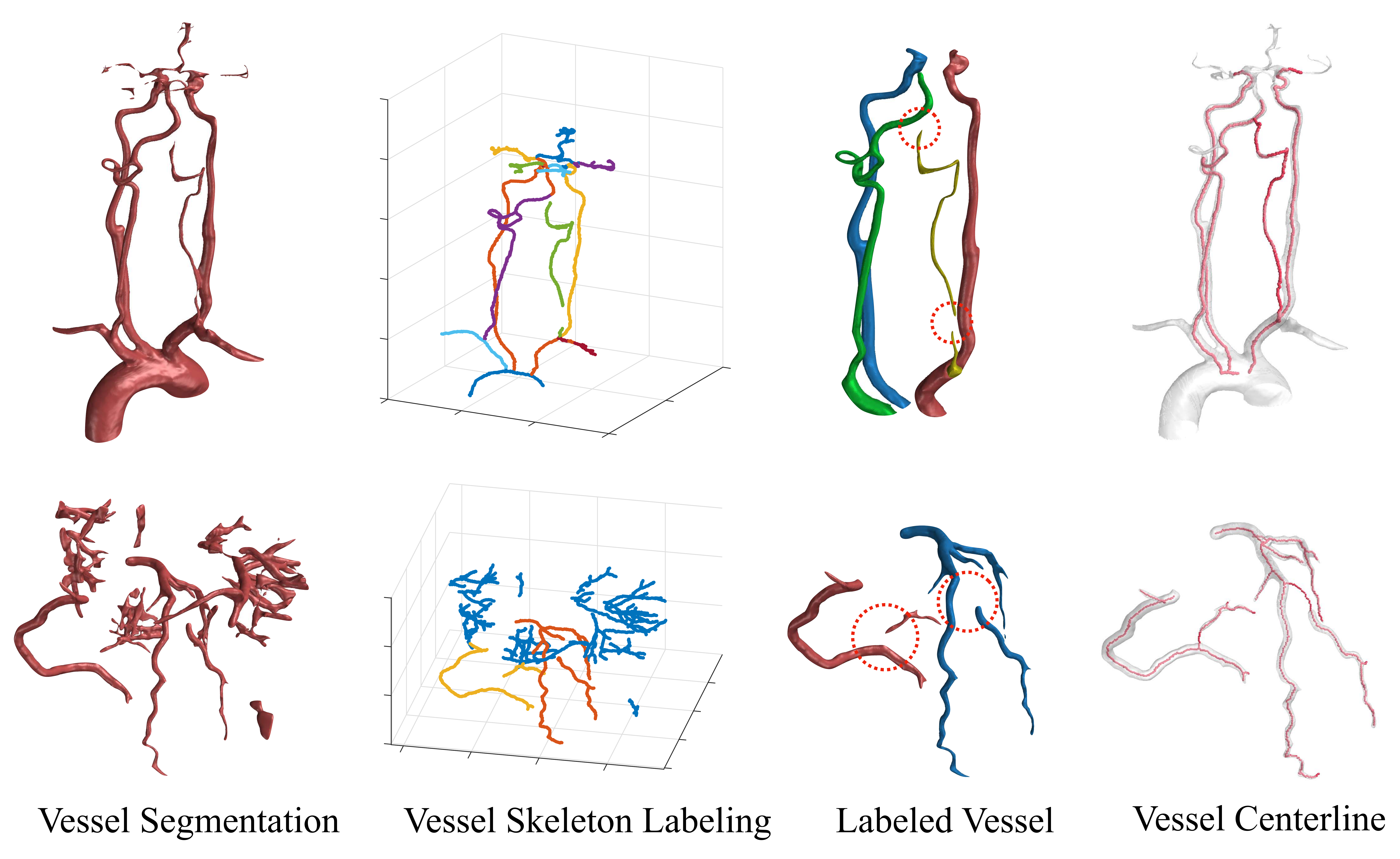}
\caption{\textbf{Vessel segmentation, labeling and centerline extraction.} \textbf{Top:} head and neck arteries. Note that the discontinuous RVA is connected in the centerline extraction procedure. \textbf{Bottom:} coronary arteries. Note that the non-vascular tissues are removed in the vessel labeling procedure and the discontinuous vessel segments are connected.}
\label{fig9}
\end{figure*}

\section{Experiments}
We evaluate the proposed method on two datasets: a public coronary artery dataset and a private head and neck artery dataset. The first dataset is mainly used to compare our method with existing baseline methods in literature. The second dataset is mainly used for ablative study to verify our system designs. Following ~\cite{schaap2009standardized}, extracted centerlines are evaluated based on three metrics, namely total overlap (OV), overlap until first error (OF), and overlap with the clinically relevant part of the vessel (OT). The stage-wise results for the proposed framework are demonstrated in Figure \ref{fig9}.

\noindent\textbf{Implementation details}
In vessel segmentation, we randomly crop 3D patches with the size of $256\times256\times32$ for the head and neck dataset and $32\times32\times32$ for the coronary artery dataset. The ResNet34 is used as the encoder in the UNet architecture which starts with 32 feature channels that are doubled in each scale, and the max-pooling layer is removed from the original residual network. All convolutions are specified as $3\times3\times3$ kernels, except the last two ResNet blocks, which are $3\times3\times1$ to reduce the parameter count. We employ the Adam optimizer with a polynomial learning rate
which equals to
$0.001\times(1-\frac{iter}{total\,iter})^{0.9}$. In vessel labeling, the inputs are skeleton points, which are generated from vessel segments and resampled to 3000 points for each sample. The inital learning rate is set to 0.001, which is reduced by half every 30 epochs. The weight $\lambda_{n}$ in GAG is set to be 0.3.

\noindent\textbf{Experiments on the Coronary Artery Dataset}
This public dataset contains 100 cardiac CT angiography (CCTA) scans collected from the clinic for training and 32 CCTA scans from ~\cite{schaap2009standardized} for evaluation. We train the vessel segmentation network and vessel labeling network on the annotated coronary artery CTA images and the vessel skeletons are labeled as three categories including right arteries, left arteries and false-positive venous vessels. According to the ablation study in the head and neck artery dataset, we use the Pointnet++ with the GAG module as our vessel labeling network.
The quantitative comparison is listed in Table \ref{table2} and the visualization of results is showed in Figure \ref{fig8}. Our hybrid approach achieves the highest performance in terms of OV and OT, respectively, indicating that the centerlines extracted by the proposed method are more complete than those produced by other methods.

\noindent\textbf{Experiments on the Head and Neck Artery Dataset.}
This private dataset collected from the clinic contains 450 CTA scans, each of which has a manually annotated vessel mask. The dataset is split into 380 scans for training, 20 for validation and 50 for testing. In the dataset, vessel skeletons are labeled as 17 categories including left and right common carotid artery (L/RCCA), left and right vertebral artery (L/RVA), etc.
Table 2 shows the evaluation results of several variants of our system with different point-cloud network designs. It can be seen that, with the geometry-aware grouping (GAG) method, the vessel labeling accuracy of both PointNet++ and DGCNN can be improved. Figure \ref{fig7} shows the GAG can facilitate local consistency of the skeleton components.

\section{Conclusions}
We propose an automatic and template-free approach to 3D vessel centerline extraction based on hybrid representations, which ensures the connectivity of extracted centerlines. We show that the hybridization between learning local appearance with patch-based CNNs and learning global geometry with point-cloud networks results in an efficient and robust framework to extract geometric objects from 3D data. We demonstrate superior performance on artery centerline extraction from CTA images and believe that the proposed approach can also be applied in other centerline or skeleton extraction tasks.

\subsubsection{Acknowledgements}
This work is funded by National Key Research and Development Program of China (No. 2019YFC0118100), and is partially supported by National Key Research and Development Program of China with Grant No. 2018AAA0101900/2018AAA0101902, Beijing Municipal Commission of Science and Technology under Grant No. Z181100008918005, the National Natural Science Foundation of China (NSFC Grant No. 61772039 and No. 91646202), and Hong Kong Research Grant Council [12301417, 16307818, 16301419]; Hong Kong University of Science and Technology [R9405, IGN17SC02, Z0428].

\bibliographystyle{splncs04}
\bibliography{paper604}

\end{document}